\documentclass[twocolumn,amsmath,amssymb,prl,groupedaddress,nobibnotes,showpacs]{revtex4}
\usepackage{graphicx}

\begin{document}

\title{Quantum Random Number Generator using Photon-Number Path Entanglement}

\author{Osung Kwon}
\affiliation{Department of Physics, Pohang University of Science and Technology (POSTECH), Pohang, 790-784, Korea}

\author{Young-Wook Cho}
\affiliation{Department of Physics, Pohang University of Science and Technology (POSTECH), Pohang, 790-784, Korea}

\author{Yoon-Ho Kim}
\email{yoonho@postech.ac.kr}
\affiliation{Department of Physics, Pohang University of Science and Technology (POSTECH), Pohang, 790-784, Korea}

\date{\today}

\begin{abstract}
We report a novel quantum random number generator based on the photon-number$-$path entangled state which is prepared via two-photon quantum interference at a beam splitter. The randomness in our scheme is of truly quantum mechanical origin as it comes from the projection measurement of the entangled two-photon state. The generated bit sequences satisfy the standard randomness test.
\end{abstract}

\pacs{03.67.-a, 03.67.Dd, 42.50.St, 42.50.Dv}

\maketitle

%%%%%%%%% Paper Body %%%%%%%%%%%

The need for generating random numbers arises in many scientific and engineering disciplines, in addition to obvious gaming industries. For instance, in quantum cryptography, the initial choices of the basis and the polarization state for the photon must be truly random for a secure system. Although mathematical algorithm may be used to obtain (pseudo-)random numbers which exhibit some statistical random behaviors, they are not truly random in the sense that the algorithmic method is deterministic in nature.  

Since randomness is inherent in quantum physics, a physical random number generator built upon a quantum mechanical process would offer true randomness. For example, consider a single-photon, $|1\rangle$, entering a lossless 50/50 beam splitter via one of the two input ports. The state at the output ports of the beam splitter is easily calculated to be in quantum superposition, $|\psi\rangle = (|1\rangle_t + |1\rangle_r)/\sqrt{2}$, where the subscripts $t$ and $r$ refer to the two output modes of the beam splitter. The single-photon detectors placed at the output ports perform the projection measurement on the quantum state $|\psi\rangle$: when the $t$ ($r$) detector clicks, we know that the quantum state has collapsed to $|1\rangle_t$ ($|1\rangle_r$). It is easy to see that each detector has 50\% probability of registering the photon but, in the framework of quantum physics, it is not possible to predict which of the two detectors will click. Since the final outcome is non-deterministic and quantum mechanically random, this process may then be used to build a quantum random number generator (QRNG). 

The above discussion, hence, allows us to identify that the key elements that give rise to quantum mechanical randomness as the \textit{projection measurement} and the \textit{quantum superposition state}. In other words, quantum mechanical randomness arises from the projection measurement on a quantum superposition state. 

Obviously, the single-photon beam splitting scheme discussed above would make the ideal QRNG if properly implemented. Unfortunately, an efficient single-photon source which is essential for the beam splitter-based QRNG does not yet exist and the scheme, in practice, is implemented with attenuated optical pulses \cite{jennewein,stefanov,wang}. Thus, practical implementations of the beam splitter-based QRNG scheme do not properly realize the key elements for the QRNG.

Recently, a number of alternative QRNG schemes have been reported in literature \cite{wu,stipcevic,wayne,ma,fiorentino}. Some of these schemes make use of Poissonian statistics inherent in the photon emission and detection processes to extract random bit sequences \cite{wu,stipcevic,wayne}. The key elements of QRNG, i.e., projection measurement and quantum superposition, however, have not been invoked in these schemes. Implementations of QRNG which do in fact realize the projection measurement and the quantum superposition state are reported in Ref.~\cite{ma,fiorentino}. In Ref.~\cite{ma}, the beam splitter-based QRNG scheme was implemented with the heralded single-photon state from spontaneous parametric down-conversion (SPDC). This implementation, however, is based on postselection of the detected events, requiring two detectors and a coincidence circuit for coincidence-based postselection \cite{hong}. In Ref.~\cite{fiorentino}, a QRNG scheme involving a two-photon polarization entangled state is reported \cite{fio2}. While this scheme could, in principle, realize the projection measurement on a quantum superposition state, preparing and maintaining a pure polarization entangled state are difficult experimental tasks and often require quantum state tomography.

In this paper, we report a novel quantum random number generator based on the photon-number$-$path entangled state which is prepared via two-photon quantum interference at a beam splitter. Since the randomness in our scheme comes from the projection measurement on the two-photon photon-number$-$path entangled state, true quantum mechanical randomness can be guaranteed for the generated bit sequences. Furthermore, the use of photon-number$-$path entanglement not only makes the pure state preparation easier, but also removes the lengthy quantum state tomography process from state characterization.

\begin{figure}[t]
\centering
\includegraphics[width=3.4in]{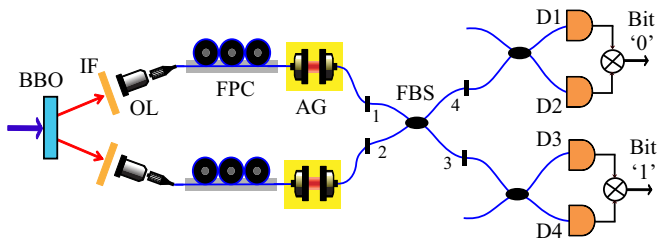}
\caption{Schematic of the experiment. The two-photon photon-number$-$path entangled state is prepared by interfering the SPDC photon pair at a beam splitter (FBS). The coincidence events between detectors D1-D2 and D3-D4 form the bit value 0 and the bit value 1, respectively.
}\label{setup01}
\end{figure}

The schematic of the quantum random number generator using the photon-number$-$path entanglement is shown in Fig.~\ref{setup01}. A pair of 816 nm photons are generated in a 1 mm thick type-II BBO crystal via the SPDC process pumped by a 408 nm diode laser. The type-II BBO crystal is phase-matched so that the signal-idler photon pairs are emerging from the crystal in the beam-like configuration at the angle of $\pm 3.2^\circ$ with respect to the pump laser \cite{beam}. The 40 mW pump laser was focused at the BBO crystal using a $f=300$ mm lens. A pair of 10 nm full-width half-maximum (FWHM) interference filters (IF) are used to cut the unwanted pump wavelength noise and the signal/idler photons are coupled to a single-mode optical fiber using a $\times 10$ objective lens. The signal-idler photons are then brought together at a 3 dB (or 50/50) fiber beam splitter (FBS). The fiber polarization controllers (FPC) are used to ensure that the arriving photons have the same polarization state and the adjustable air-gaps (AG) allow us to control the arrival time difference between the photons at FBS. 

%%%%%%%%%%%%%%%%%%
%%%%%%%%%%%%%%%%%

The input quantum state to the FBS is, thus, written as $|\psi\rangle_i = |1\rangle_1 \otimes |1\rangle_2$, referring to the two photon state where each photon of the SPDC photon pair occupying one of the two input modes of the FBS. It is well-known that when a pair of identical photons arrive simultaneously at a 50/50 beam splitter via different input ports, two-photon quantum interference takes places and, as a result, the two photons always exit the beam splitter via the same output port \cite{hom,nooverlap}. The quantum state occupying the two output modes of the FBS is then calculated to be,
%%%%%
\begin{equation}\label{2002}
|\psi\rangle =\left(|2\rangle_3\otimes|0\rangle_4 + |0\rangle_3\otimes|2\rangle_4 \right)/\sqrt{2},
\end{equation}
%%%%%
where, for example, $|2\rangle_3\otimes|0\rangle_4$ refers to the probability amplitude of finding two photons (zero photons) in the lower (upper) output mode of the FBS. 

Since the preparation of the entangled state in eq.~(\ref{2002}) is at the heart of the present QRNG scheme, it is of utmost importance to experimentally  prepare a high-purity photon-number$-$path entangled state. To verify the preparation of eq.~(\ref{2002}), we can make use of the two-photon ``dip'' by connecting the two output fibers  (modes 3 and 4) of the FBS to single-photon counting detectors and observing the coincidence count between them.  As demonstrated in Ref.~\cite{hom}, high visibility (approaching 100\%) two-photon dip is the de facto signature of the two-photon photon-number$-$path entangled state in eq.~(\ref{2002}).

%%%%%%%%%%%%%%%%%%
\begin{figure}[t]
\centering
\includegraphics[width=3.4in]{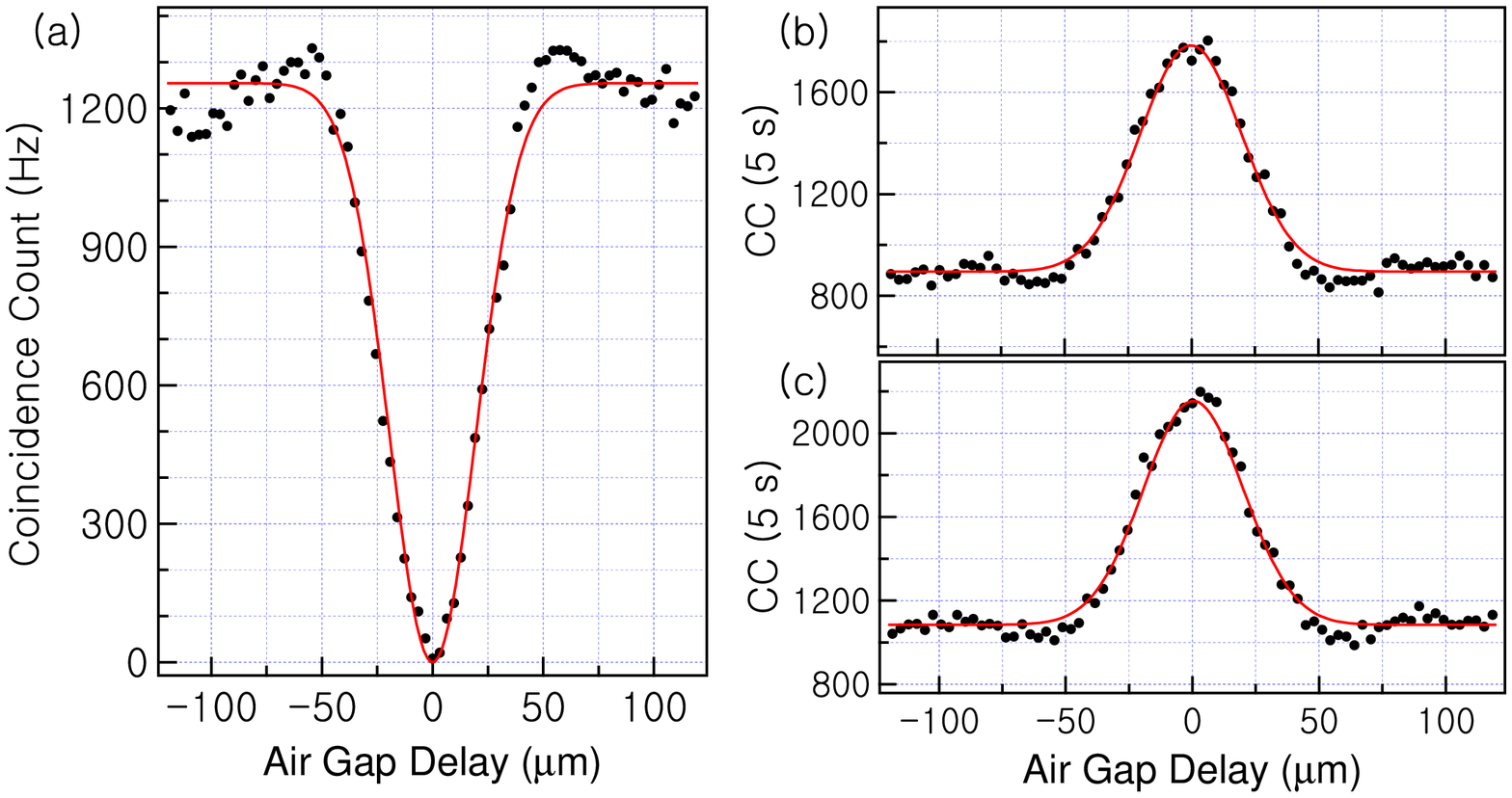}
\caption{(a) The coincidence between two detectors placed at the end of the first FBS. The solid line is a Gaussian fit to the data with 100\% visibility. D1-D3 coincidence shows a dip with the same visibility. (b) D1-D2 coincidence and (c) D3-D4 coincidence exhibit peaks at the delay where the dip occurs.
}\label{data01}
\end{figure}
%%%%%%%%%%%%%%%%%

Figure \ref{data01}(a) shows the experimental data, i.e., coincidences between single-photon detectors connected to modes 3 and 4 of FBS. The data demonstrates near-perfect two-photon visibility: the solid line is a Gaussian fit to the data with 100\% visibility. The coincidence between detectors D1 and D3 shows a dip with the same visibility. We note that, in the present QRNG, observing the two-photon dip is sufficient to characterize the quantum state needed for the QRNG operation. This process is much simpler and easier than characterizing the two-photon polarization entanglement via quantum state tomography \cite{fiorentino}.

The projection measurement on eq.~(\ref{2002}) would then reveal quantum mechanical randomness and this may be accomplished by connecting a photon number resolving detector at each output mode of the FBS. Since we only need to resolve the two-photon state, $|2\rangle$, we implemented the photon number resolving detector with a fiber beam splitter (FBS), two single-photon counting detectors, and a coincidence circuit (with a 3 ns coincidence window), as shown in Fig.~\ref{setup01}. Provided that the quantum state in eq.~(\ref{2002}) is being measured, the coincidence event between D1-D2 or D3-D4 detectors tells us that the state has collapsed to $|0\rangle_3\otimes|2\rangle_4$ or $|2\rangle_3\otimes|0\rangle_4$, respectively. Figures \ref{data01}(b) and \ref{data01}(c) show the D1-D2 and D3-D4 coincidence measurements, respectively, as a function of the air gap delay: the peaks at the zero air gap delay are the result of the projection measurement on the photon-number$-$path entangled state in eq.~(\ref{2002}) \cite{kim03a}.

Clearly, the projection measurement on eq.~(\ref{2002}) would leave us either with a D1-D2 coincidence or a D3-D4 coincidence and this event is truly quantum mechanically random with equal probability. We can thus make use of these coincidence events to generate a bit sequence which is truly quantum mechanically random. We emphasize that for this type of coincidence measurement to have the above operational interpretation, it is essential to ensure that what is being measured is the photon-number$-$path entangled state in eq.~(\ref{2002}). (To do this, we keep the air gap at the zero delay position and constantly monitor the D1-D3 coincidence.) Otherwise, the output becomes a classical random process due to multiple beam splitters and coincidence measurements.

\begin{figure}[t]
\centering
\includegraphics[width=3.4in]{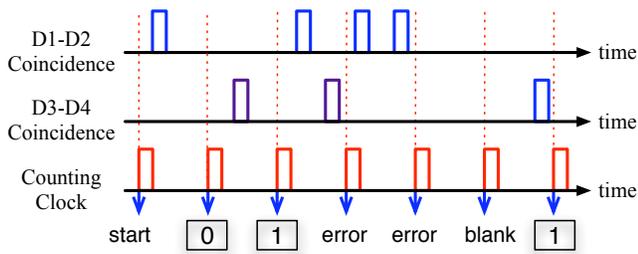}
\caption{Bit sequence generation scheme. D1-D2 coincidence and D3-D4 coincidence occur randomly due to the  entangled state in eq.~(\ref{2002}). If there is a D1-D2 or D3-D4 coincidence event between two successive counting clocks pulses, we record the bit value 0 or 1, respectively. If there are two or more such events within the time period, we record that as an error. The error bit can be removed by increasing the counting clock frequency. 
 }\label{setup02}
\end{figure}

The schematic for the random bit sequence generation is shown in Fig.~\ref{setup02}. A counter/timer board (National Instruments PCI-6602) records the bit sequences using an external function generator as the counting clock. Whenever there is a coincidence event  (D1-D2 or D3-D4) between the two adjacent clock pulses, the event is recorded as the bit value 0 for the D1-D2 coincidence and the bit value 1 for the D3-D4 coincidence. For the case where there are two or more events within the time period, we record an error bit at the next clock pulse. Note that all the error bits in our scheme originate from these multiple D1-D2 or D3-D4 coincidence events during the clock sequence because D1-D2-D3-D4 four-fold coincidence does not occur. In other words, D1-D3 coincidence shows no counts while the QRNG experiment is being performed at the zero air gap delay.

%%%%%%%%%%%%%%%%%%
%%%%%%%%%%%%%%%%%

\begin{figure}[t]
\centering
\includegraphics[width=3.4in]{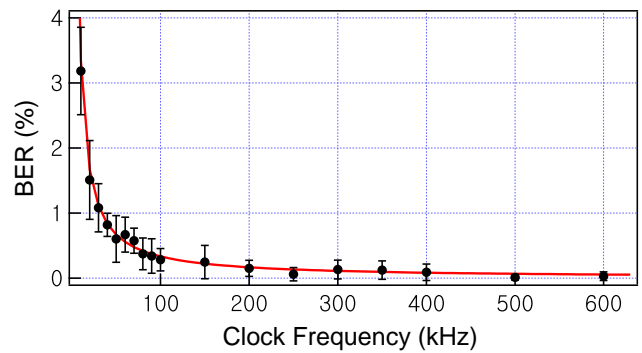}
\caption{BER versus the counting clock frequency. Solid line is eq.~(\ref{ber}) with $R_B=668$ Hz. Solid circles are the experimental data with one standard deviation error bars. }\label{data02}
\end{figure}

The bit error rate (BER), the ratio of the number of error bits to the number of total bits, can be evaluated as follows. Assuming that the probability of a single D1-D2 or D3-D4 coincidence event during the clock cycle is equal, the probability of generating a bit is given as $(1-t)/f$, where $f$ is the clock frequency and the $t$ varies between $0\le t \le 1$. Then, the probability of this bit to become an error bit is $\int_0^1 R_B (1-t)/f dt = R_B/2f$, where $R_B$ is the bit generation rate determined from the coincidence rates D1-D2 and D3-D4. If $N$ is the number of total bits in the sequence, the number of error bits is $N R_B/2f$. The BER thus is calculated to be,
%%%%%
\begin{equation}\label{ber}
\textrm{BER} =  {R_B}/{2 f}.
\end{equation}
%%%%%%%
The above relation was tested by measuring the BER versus the clock frequency. In this experiment, $R_B=668$ Hz, determined from the coincidence rates. Fig.~\ref{data02} shows the experimental data and they are in good agreement with eq.~(\ref{ber}). 

The random bit sequences are then recored at the clock frequency of 500 kHz where the experimental BER is zero: higher clock rates do not improve the random bit generation rate. To make sure the two-photon state in eq.~(\ref{2002}) is maintained for the duration of the random bit recording session, D1-D3 coincidence was monitored at all times: non-zero D1-D3 coincidence events are not present in our experimental data.

 %%%%%%%%%%%%%%%%%%
%%%%%%%%%%%%%%%%%

The recorded random binary sequence of 0's and 1's should in principle be unbiased, i.e., the numbers of 0's and 1's should be the same. However, due to experimental imperfections, such as, detection efficiency mismatches, different optical losses, etc., the 0's occur somewhat less than the 1's in our experiment as shown in Figs.~\ref{data01}(b) and \ref{data01}(c). We thus applied the well-known unbiasing algorithm to the recorded bit sequence: two successive bits are grouped together, forming four possible pairs 00, 01, 10, and 11. If the binary sequence is biased, the probabilities of 00 and 11 are not equal so we discard these groups. The probabilities of 01 and 10 are, however, equal so we convert the bit group 01 as 0 and 10 as 1  \cite{neumann,peres}. As a result, we are left with a set of unbiased random binary sequence of 0's and 1's. In our experimental data, the length of the final unbiased bit sequence was 23.96\% of that of the original biased bit sequence.

%%%%%%%%%%%%%%%%%%%%%%%%%%%
\begin{table}[t]
\caption{\label{table} Results ($P$-$values$) of the statistical randomness test by using NIST STS \cite{nist}. Three sets of 1 Mbit-long unbiased random binary sequences generated by the QRNG were tested. These results indicate that there are no statistical patterns in the tested random number sets.  }
\begin{ruledtabular}
\begin{tabular}{lccccr}
 & Test & Set 1 & Set 2 & Set 3 & \\
\hline
 & Approx. Entropy & 0.03385 & 0.89904 & 0.05049 &\\
 & Block Frequency & 0.02526 & 0.59674 & 0.04524 &\\
 & Cumulative-sums & 0.24325 & 0.68146 & 0.44076 &\\
 & FFT & 0.10392 & 0.31056 & 0.31494 &\\
 & Frequency & 0.22990 & 0.94286 & 0.36920 &\\
 & Linear Complexity & 0.23791 & 0.13744 & 0.55239 &\\
 & Longest Run & 0.23024 & 0.54439 & 0.38463 &\\
 & Non-overlapping Templates & 0.53029 & 0.54345 & 0.51499 &\\
 & Overlapping Template & 0.80689 & 0.66523 & 0.49260 &\\
 & Rank & 0.97541 & 0.34506 & 0.91025 &\\
 & Random Excursions & 0.56713 & 0.56957 & 0.76068  &\\
 & Random Excursions variant & 0.39073 & 0.52697 & 0.74450 &\\
 & Runs & 0.92430 & 0.13432 & 0.51448 &\\
 & Serial & 0.24637 & 0.32450 & 0.26694 &\\
 & Universal & 0.40244 & 0.79281 & 0.75569 &\\
\end{tabular}
\end{ruledtabular}
\end{table}
%%%%%%%%%%%%%%%%%%%%%%%%%%%%%%%%%%

Three sets of 1 Mbit-long (1,090,000 to be exact) unbiased random binary sequences are recorded and the randomness of the sequences are tested by using the widely used NIST statistical test suite (STS) \cite{nist}. The STS consists of a set of 15 randomness tests, evaluating the \textit{P-value} to quantify the randomness of the sequence. In short, if the \textit{P-value} is smaller than the significance level $\alpha$, the STS concludes that the sequence is not random with a confidence of $1-\alpha$. The significance level of 0.01 (the default setting of the STS and a common value used in cryptography) was chosen for the tests and the three random sequences passed all of the tests in the STS, see Table \ref{table}. We note that the STS test does not guarantee randomness unless the length of the sequence is infinitely long \cite{nist}. The results, however, indicate the absence of any statistical patterns in the sequence.

In summary, we demonstrated a novel quantum random number generator based on the two-photon photon-number$-$path entangled state. Since the randomness in our scheme is based on the projection measurement of the entangled two-photon state, the bit sequence is truly quantum mechanically random. In addition, our QRNG scheme is simple to implement and characterize compared to other schemes. Finally, we note that the random bit generation rate can be substantially improved by utilizing a high-brightness two-photon source based on quasi-phase-matched crystals \cite{zeilinger}.

This work was supported, in part, by the Korea Science and Engineering Foundation (R01-2006-000-10354-0),  the Korea Research Foundation (KRF-2006-312-C00551), and by the Ministry of Commerce, Industry and Energy of Korea through the Industrial Technology Infrastructure Building Program.

%%%%%%%%%%%%%%%%%%%%%%%%%%%%%%%%%%%%%%
%\begin{thebibliography}{}


\begin{thebibliography}{99}
\bibitem{fio2} The implementation of the beam splitter QRNG scheme reported in Ref.~\cite{fiorentino} makes use of only the signal photon of the SPDC signal-idler photon pair. The scheme, therefore, is equivalent to illuminating a beam splitter with a weak thermal light \cite{stre,baek08}.
\end{thebibliography}
\end{document}